\documentclass[12pt]{iopart}

\usepackage{graphicx}
\usepackage{epstopdf}
\usepackage{upgreek}
\usepackage{citesort}
\usepackage{todonotes}
\usepackage{placeins}

\newcommand{\lk}{L_\mathrm{k}}
\newcommand{\kb}{k_\mathrm{B}}
\newcommand{\ic}{I_\mathrm{c}}
\newcommand{\isw}{I_\mathrm{sw}}
\newcommand{\std}{\sigma_\mathrm{sw}}
\newcommand{\mean}{\langle I_\mathrm{sw}\rangle}
\newcommand{\tc}{T_\mathrm{c}}
\newcommand{\bl}{\beta_\mathrm{L}}
\newcommand{\lloop}{L_\mathrm{loop}}

\begin{document}

\title[Onset of phase diffusion in high kinetic inductance granular aluminum micro-SQUIDs]{Onset of phase diffusion in high kinetic inductance granular aluminum micro-SQUIDs}

\author{Felix Friedrich$^1$\footnote{Present address:
Physikalisches Institut, Experimentelle Physik II, Universit\"at W\"urzburg, Am Hubland, 97074 W\"urzburg, Germany}, Patrick Winkel$^1$, Kiril Borisov$^2$, Hannes Seeger$^1$, Christoph S\"urgers$^1$, Ioan M. Pop$^{1,2}$, Wolfgang Wernsdorfer$^{1,2,3}$}

\address{$^1$ Physikalisches Institut, Karlsruhe Institute of Technology, Wolfgang-Gaede-Str.~1, 76131 Karlsruhe, Germany}
\address{$^2$ Institute of Nanotechnology, Karlsruhe Institute of Technology, Hermann-von-Helmholtz-Platz 1, 76344 Eggenstein-Leopoldshafen, Germany}
\address{$^3$ Institut N\'{e}el, CNRS and Universit\'{e} Grenoble Alpes, BP 166, 25 Avenue des Martyrs, Grenoble 38042, Cedex 9, France}
\eads{\mailto{felix.friedrich@physik.uni-wuerzburg.de}, \mailto{wolfgang.wernsdorfer@kit.edu}}

\vspace{10pt}

\begin{abstract}
Superconducting granular aluminum is attracting increasing interest due to its high kinetic inductance and low dissipation, favoring its use in kinetic inductance particle detectors, superconducting resonators or quantum bits. We perform switching current measurements on DC-SQUIDs, obtained by introducing two identical geometric constrictions in granular aluminum rings of various normal-state resistivities in the range from $\rho_\mathrm{n}=250\,\upmu\Omega\,\mathrm{cm}$ to $5550\,\upmu\Omega\,\mathrm{cm}$. The relative high kinetic inductance of the SQUID loop, in the range of tens of nH, leads to a suppression of the modulation in the measured switching current versus magnetic flux, accompanied by a distortion towards a triangular shape. We observe a change in the temperature dependence of the switching current histograms with increasing normal-state film resistivity. This behavior suggests the onset of a diffusive motion of the superconducting phase across the constrictions in the two-dimensional washboard potential of the SQUIDs, which could be caused by a change of the local electromagnetic environment of films with increasing normal-state resistivities.

\end{abstract}

%
\vspace{2pc}

%
\maketitle
%
%

\section{Introduction}

Superconductivity in granular aluminum (grAl) films was first reported by Cohen and Abeles in 1968~\cite{Cohen1968}. The material consists of crystalline nanometer-sized aluminum (Al) grains embedded in an amorphous aluminum oxide matrix; the grains form when Al is evaporated in an oxygen (O$_2$) atmosphere~\cite{Cohen1968,Deutscher1973}. Granular aluminum was originally studied due to its increased critical temperature of up to 3\,K~\cite{Deutscher1973} and its critical field of more than 3\,T~\cite{Chui1981_2}. A detailed study on the origin of the increased critical temperature of grAl was recently presented by Pracht~\textit{et al.}~\cite{Pracht2016}. Transport measurements in the normal state of grAl thin films demonstrated Kondo-like behavior of the film resistivity~\cite{Bachar2013}, and modeling superconducting grAl as an array of Josephson junctions showed good agreement with experiments on superconducting microwave resonators~\cite{Maleeva2018}. 

From an application point of view, grAl is interesting due to the fact that it can exhibit large kinetic inductance values and low losses in the microwave domain~\cite{Sun2012,Rotzinger2017,Grunhaupt2018}. The susceptibility of the kinetic inductance to temperature, changes in the superfluid density, and bias current suggests the use of grAl elements in kinetic inductance detectors~\cite{Day2003,Valenti2018} and tunable microwave resonators~\cite{Vissers2015}. Furthermore, compact high impedance components are needed for the implementation of advanced qubit designs~\cite{Manucharyan2009,Pop2010,Grunhaupt2018_2}.

Here we present switching current measurements on grAl DC-SQUIDs (direct current superconducting quantum interference devices), which can be complementary to radio-frequency (rf) measurements. Unlike rf measurements, which provide information in a narrow frequency band in the vicinity of the eigenmodes of the measured structures, switching current measurements are sensitive to a broad spectrum of frequencies. Concretely, the switching dynamics of the measured SQUIDs can indicate changes at the high end of the spectrum close to the superconducting gap~\cite{Tinkham1996}, where it is difficult to perform accurate rf experiments. As we will show in section~\ref{sec:results}, our main result is the observation of a change in the temperature dependence of the switching current histogram width~$\std$. For standard Al SQUIDs, $\std$ increases with temperature, as expected~\cite{Li2011}, while for high impedance grAl SQUIDs $\std$ decreases with temperature. This indicates the onset of phase diffusion~\cite{Krasnov2005,Kivioja2005,Mannik2005}, which could be linked to additional damping at frequencies comparable to the plasma frequency~$\omega_\mathrm{p}$~\cite{Tinkham1996}.

\section{Experimental}
\subsection{Sample fabrication}
The measured micro-SQUIDs consist of a superconducting loop with an area of 1\,$\upmu\mathrm{m}^2$, interrupted by two identical geometric constrictions with a length $l\sim300$\,nm and width $w\sim80$\,nm. An SEM image of a typical sample is shown in the inset of \fref{fig:1}.  The layout is patterned by electron-beam lithography into a double layer PMMA resist stack on a degenerately p-doped Si/SiO$_2$ wafer. The grAl and Al films are deposited by electron beam evaporation at ambient substrate temperature, with thicknesses between 20\,nm and 30\,nm.  The grAl normal state resistivity is tuned by adjusting the partial pressure of O$_2$ in the chamber. The resulting normal state resistivities of the three grAl samples measured at room temperature are 250\,$\upmu\Omega\,\mathrm{cm}$, 3200\,$\upmu\Omega\,\mathrm{cm}$ and 5550\,$\upmu\Omega\,\mathrm{cm}$.

In order to get an estimate of the coherence length~$\xi$ of our films, we describe grAl as a superconductor in the dirty limit~\cite{Likharev1979}. We derive the electron mean free path~$l_\mathrm{F}$ from the resistivity~\cite{Cohen1968} and use the coherence length $\xi_0=1.6\,\upmu$m of pure Al~\cite{deGennes1966} to calculate the effective coherence length \mbox{$\xi\approx 0.85\sqrt{\xi_0l_\mathrm{F}}$}~\cite{Tinkham1996} of our grAl films. The results are listed in table~\ref{tab:lengths} for the three grAl films, together with the normal state resistivity~$\rho_\mathrm{n}$, the London penetration depth~$\lambda_\mathrm{L}$ and the plasma frequency~$\omega_\mathrm{p}$ of the films. Note that $\xi$ approaches the grain size for the highest resistive samples, which is the limit for an array of Josephson junctions formed by the Al grains and the surrounding aluminum oxide matrix~\cite{Maleeva2018}. Moreover, the coherence length is smaller than the width of the geometric constrictions for all films, which means they do not form so-called constriction weak links~\cite{Tinkham1996}, but rather should be viewed as an array of effective Josephson junctions with lower critical current compared to the rest of the SQUID loop. The London penetration depth~$\lambda_\mathrm{L}$ of the grAl films, also derived from the normal state resistivity of the samples~\cite{Barone1982}, is on the scale of a few micrometers and much larger than the width and thickness of the grAl circuit traces, implying a homogeneous current density.

\begin{table}
\caption{\label{tab:lengths}Coherence lenght~$\xi$, London penetration depth~$\lambda_\mathrm{L}$ and plasma frequency $\omega_\mathrm{p}$ of the presented grAl samples. The coherence length and penetration depth are derived from the measured resistivities; $\xi$ is obtained using the coherence length of a dirty superconductor and is only a rough approximation. For large resistivities, it approaches the grain size. The plasma frequency of a grAl film similar to sample grAl~B was measured in~\cite{Maleeva2018}, the other two values are estimated from the results in the same reference.} 

\begin{indented}
\lineup
\item[]\begin{tabular}{@{}*{5}{l}}
\br                              
Sample & $\rho_\mathrm{n}$ ($\upmu\Omega$\,cm) & $\xi$ (nm) & $\lambda_\mathrm{L}$ ($\upmu$m) & $\omega_\mathrm{p}/2\pi$ (GHz)\cr 
\mr
grAl A & $\0250\pm \020$ & $27.2\pm1.1$ & $1.11\pm 0.04$ & $300$\cr 
grAl B & $3200\pm 400$ & $\07.6\pm0.7$ & $4.0\0\pm 0.2$  & $\070$\cr
grAl C & $5550\pm \090$ & $\05.8\pm0.1$ & $5.54\pm 0.04$ & $\055$\cr 
\br
\end{tabular}
\end{indented}
\end{table}

\subsection{Switching current measurements}
When biasing a DC-SQUID with a constantly increasing current, it switches to the resistive state before the critical current~$\ic$ is reached. This behavior can be understood as arising from the motion of a phase particle in the two dimensional SQUID potential~\cite{Lefevre_Seguin1992}, which is equivalent to the resistively and capacitively shunted junction model of a single Josephson junction. As the height of the potential barrier that separates the metastable states of the phase particle decreases with increasing bias current, thermal activation (TA) over the barrier or macroscopic quantum tunneling (MQT) through it will trigger switching events at bias currents smaller than $\ic$. This stochastic escape results in a distribution of switching currents~$\isw$ with defined mean value and standard deviation. Switching current distributions have already been extensively studied in various experiments~\cite{Fulton1974,Clarke1988,Wallraff2003} and show good agreement with theoretical descriptions~\cite{Kurkijarvi1972,Garg1995} of the escape of a particle over a bias dependent barrier.

We investigate the magnetic field modulation of the SQUIDs' switching currents to assess the influence of the kinetic inductance in the loop (see \fref{fig:1} and \fref{fig:2}). To examine the dynamics of the phase across the junctions in the SQUID, we present measurements of switching current distributions at different temperatures, starting from the base temperature of 20\,mK of our dilution refrigerator (see \fref{fig:3}).

The measurements are controlled by an ADwin Gold II real-time system~\cite{adwin}. Switching currents are determined by ramping the current bias until a jump in the measured voltage is detected. Before repeating the measurement, the system is allowed to thermalize. The filtering system used in our experiments was previously presented in the supplementary material of Ref.~\cite{Cleuziou2006}.

\section{Results and Discussion}\label{sec:results}

\subsection{SQUID modulation}\label{ssec:SQUIDModulation}
\begin{figure}
\begin{center}
\includegraphics[scale=1]{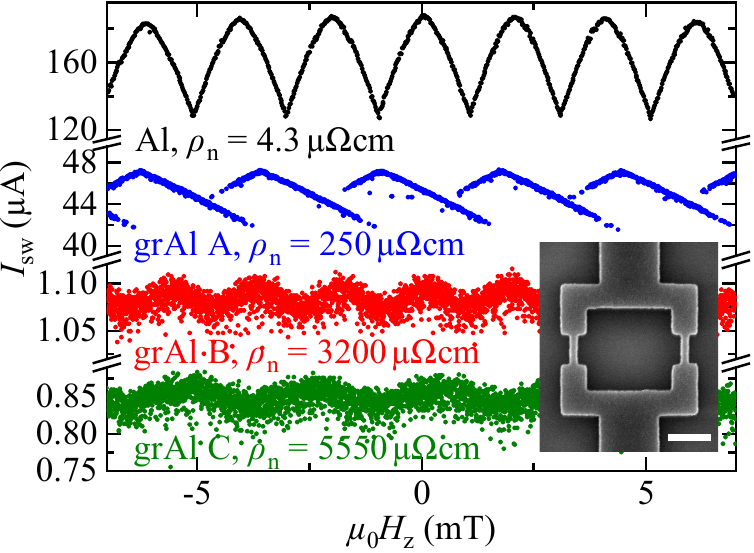}
\caption{Switching current modulation of the Al and the grAl SQUIDs with applied magnetic field. The modulation of grAl A is offset from zero due to trapped flux. Please note the breaks and different scales on the y-axis. The inset shows a scanning electron microscope image of sample grAl B. The scale bar represents a 400\,nm length. The design of all samples is identical.}
\label{fig:1}
\end{center}
\end{figure}
We measured the modulation of the switching current as a function of the magnetic field~$H_\mathrm{z}$ applied perpendicular to the SQUID plane at 20\,mK for all samples. The modulation curves are shown in \fref{fig:1}. The Al SQUID shows an almost cosine modulation with a period of approximately 2\,mT, which is in good agreement  with the estimated loop area of 1\,$\upmu\mathrm{m}^2$. The cosine modulation is expected for an Al SQUID with micro-bridge junctions, as the loop inductance is small compared to the junction inductance and the coherence length in pure Al thin films is larger than the width and the length of the constrictions. In this case, the current phase relation does not differ too much from the sinusoidal shape in tunnel junctions~\cite{Hasselbach2002}, resulting in a similar modulation of the switching current.

In comparison to the Al SQUID, the modulation curves of the grAl micro-SQUIDs show a pronounced triangular shape with a much smaller relative modulation depth. The asymmetric modulation is attributed to non-symmetric cooling after a switching event. For the grAl samples, the definition of the junction and of the corresponding current phase relation is much more subtle. As the coherence length of all samples (see table~\ref{tab:lengths}) is much shorter than the length of the constrictions, one cannot strictly speak of the constrictions as Josephson junctions~\cite{Likharev1979}. Instead, the description of grAl as an effective array of tunnel junctions suggests that the weakest junction (or junctions) in each constriction dominate the switching behavior. 

\begin{figure}
\begin{center}
\includegraphics[scale=1]{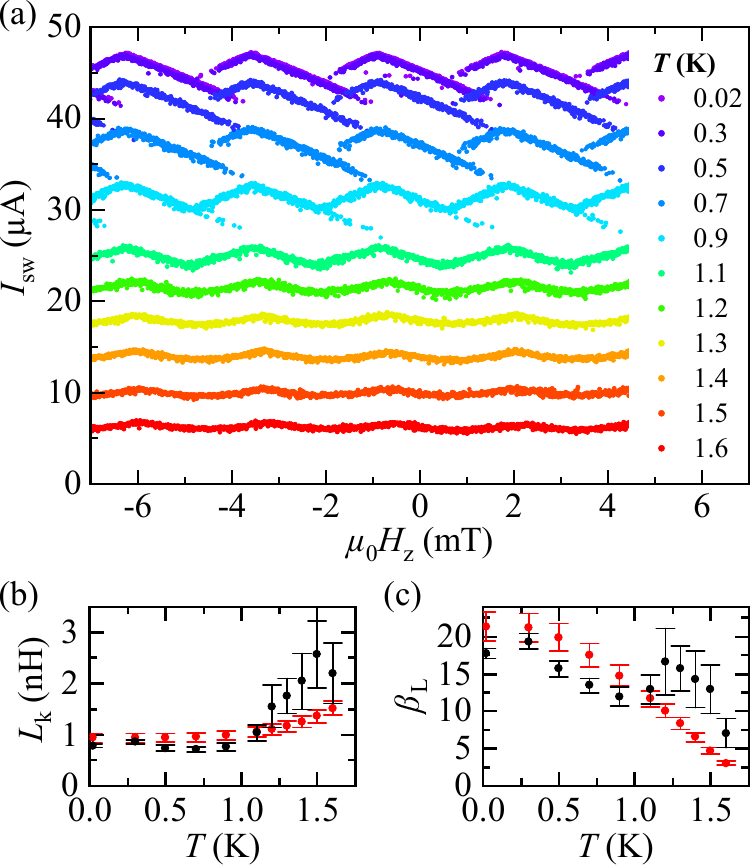}
\caption{(a)~Modulation of the switching current of sample grAl~A measured at different temperatures. Switching events from excited flux states and the triangular shape of the modulation are only visible at temperatures below 1.1\,K. (b)~Loop inductance of the grAl SQUID. Black dots are extracted from the modulation in (a) and equation\,\eref{eq:beta}, red dots are calculated from equation\,\eref{eq:lkin}. (c)~Screening factor derived from the data in (b).}
\label{fig:2}
\end{center}
\end{figure}

For a Josephson junction array with the Josephson energy dominating over the charging energy, one expects a saw-tooth like current-phase relation~\cite{Pop2010,Pop2008}. However, the influence of the loop inductance~$\lloop$ alone is sufficient to explain the triangular shape and the reduced modulation observed in our experiments~\cite{Faucher2002}. We measure the SQUID modulation of sample grAl~A at different temperatures between 20\,mK and 1.6\,K. The results are shown in \fref{fig:2}. For SQUIDs with large loop inductance, the slope in the triangular modulation is determined by the screening factor \mbox{$\bl = \ic^\mathrm{max}\lloop/\Phi_0$}~\cite{Faucher2002}, where $\ic^\mathrm{max}$ is the maximum critical current in the modulation. This directly relates the modulation depth~$\Delta \ic$ to $\bl$~\cite{Granata2016a}:
\begin{eqnarray}
\frac{\Delta \ic}{\ic^\mathrm{max}}=\frac{1}{1+\bl}.\label{eq:beta}
\end{eqnarray}
For junctions with critical currents in the range of tens of $\upmu\mathrm{A}$ or more, the measured switching currents will be very close to the critical current. Thus, using the relation in \eref{eq:beta}, we can calculate the loop inductance of the SQUID from the measured curves in \fref{fig:2}~(a) and compare it to the inductance estimated from the geometry and normal-state sheet resistance of the SQUID loop. As the geometric inductance~\cite{Jaycox1981} of our SQUIDs is small compared to the kinetic inductance, its contribution to $\lloop$ is neglected. The kinetic inductance of a ring that consists of $N_\mathrm{sq}$ sheets with a sheet resistance~$R_\mathrm{sq}$ is given by~\cite{Annunziata2010}
\begin{eqnarray}
\lk = N_\mathrm{sq}\frac{\hbar  R_\mathrm{sq}}{\pi\Delta(T)}\tanh^{-1}\left(\frac{\Delta(T)}{2\kb T}\right),\label{eq:lkin}
\end{eqnarray}
where $\Delta(T)$ is the superconducting gap parameter at temperature~$T$. $N_\mathrm{sq}$ includes the sheets in the constrictions, as they also contribute to the total inductance of the SQUID loop. Using the BCS-temperature dependence of the gap~\cite{Carrington2003}, $N_\mathrm{sq}=18\pm1$ extracted from a scanning electron microscope image of the sample, and the measured values of $R_\mathrm{sq}=(83\pm6)\,\Omega$ and $T_\mathrm{c}=(2.2\pm0.1)\,\mathrm{K}$, we can calculate $\lk$ for all temperatures investigated. In \fref{fig:2}~(b), this value is compared to $\lloop$ extracted from the relative modulation depth. We find good agreement between the calculated kinetic loop inductance and the value extracted from the measured modulation curves up to 1.1\,K, above which the modulation vanishes. For samples grAl B and C, the uncertainty of the loop inductance deduced from the modulation is large due to the wide histogram in comparison to the modulation. Still, the agreement between the loop inductance and the calculated kinetic inductance is within one order of magnitude (not shown).

Switching events below the main modulation are observed for sample grAl A. In the two-dimensional SQUID potential, these events are explained by escapes from excited states, that exist for $\bl>1$~\cite{Lefevre_Seguin1992}, which is the case for all of our samples (for sample grAl A cf. \fref{fig:2}~(c), others not shown). In the experiment, this corresponds to additional flux quanta trapped in the SQUID ring, resulting in a larger circulating current and therefore lower switching current. For sample grAl A such switching events are detected only below 900 mK. No switching events from excited states were observed for samples grAl~B and grAl~C. This indicates the existence of two different switching dynamics in the grAl junctions, which also appear in the measured switching current distributions, discussed in the following.

\subsection{Switching current distributions}
\begin{figure*}
\begin{center}
\includegraphics[scale=1]{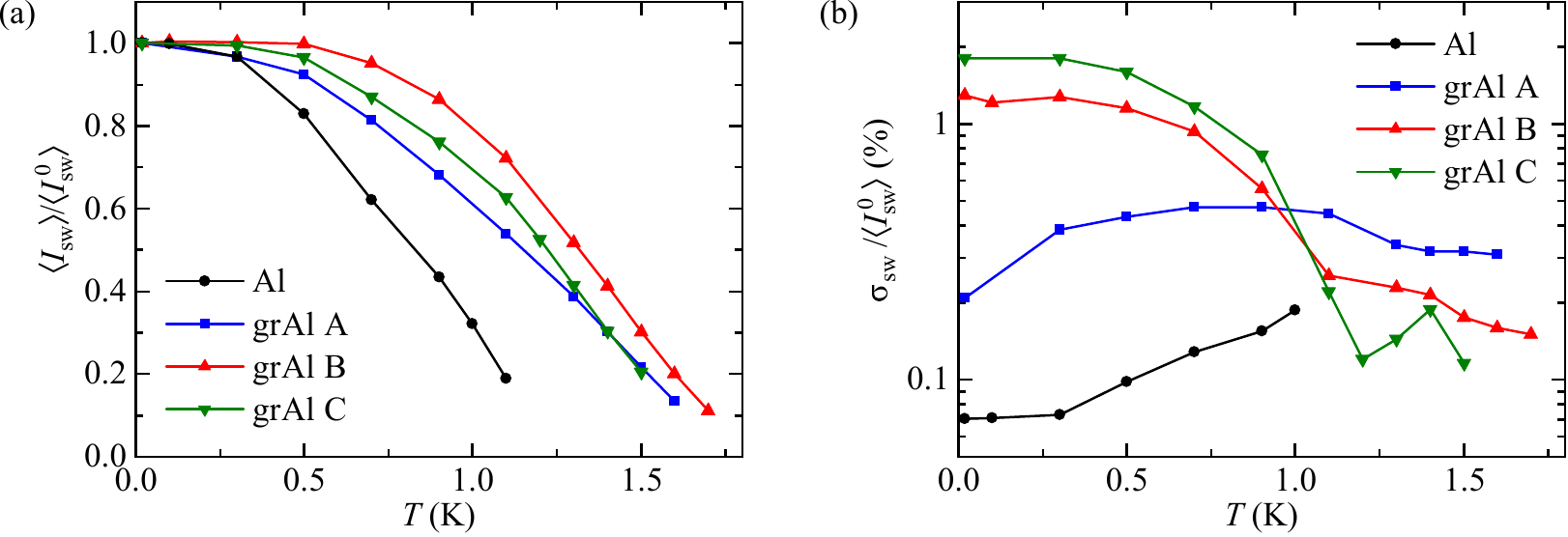}
\caption{(a) Mean value~$\mean$ of the measured switching current distributions normalized to $\mean$ measured at 20\,mK. Different current sweep rates were used for each sample: 0.5\,mA/s for Al, 186\,mA/s for grAl A, and 15\,$\upmu$A/s for grAl B and grAl C.  (b) Normalized standard deviation~$\std$ corresponding to the measurements in (a). For the Al SQUID $\std$ increases with rising temperature, while $\std$ decreases for the high resistivity grAl B and grAl C SQUID. The low resistivity grAl A SQUID shows a crossover behavior. In order to obtain sufficient statistics on the switching current distributions, at least 3000 measurements are recorded at each temperature.}
\label{fig:3}
\end{center}
\end{figure*}
The mean value~$\mean$ and the standard deviation~$\std$ of the measured switching current distributions as a function of temperature are shown in \fref{fig:3} for all four samples. For samples with relatively large critical currents (Al, grAl~A), the measured mean switching currents are expected to closely follow the temperature dependence of the critical current, due to the large potential barrier in the SQUID potential, which scales with $\ic$. For a SQUID with a maximum switching current of about 1\,$\upmu$A (grAl~B, C), $\mean$ is expected to deviate from $\ic$ significantly~\cite{Garg1995}. However, no pronounced qualitative difference between the normalized curves is visible, apart from the larger critical temperature~$\tc$ of the grAl thin films compared to pure Al. The relative width of the distributions~\mbox{$\std/\langle\isw^0\rangle$}, in contrast, differs significantly for all samples and its temperature dependence gives insight into the phase dynamics in the micro-SQUIDs. The Al SQUID shows the expected temperature dependence of $\std$ for switching events that are triggered by a single escape of the phase particle across the potential barrier: at low temperatures, temperature independent MQT of the phase particle through the barrier dominates the escape, resulting in a constant width of the switching current up to 300\,mK. Above this temperature, the contribution of TA to the escape events surpasses that of MQT and the width increases due to the increasing thermal fluctuations. In the regime of TA, the temperature dependence of the switching current distribution is given by
\begin{eqnarray}
\std\propto T^{1/\alpha}\ic(T)^{1-1/\alpha},\label{eq:width}
\end{eqnarray}
with $\alpha$ depending on the current-phase relation of the Josephson junctions~\cite{Li2011} \mbox{($\alpha=3/2$} for a sinusoidal current-phase relation). At temperatures close to the critical temperature, the increase of $\std$ is larger than expected from \eref{eq:width}, by up to a factor of two. We attribute this deviation to small temperature fluctuations of less than 1\,mK in our experiment, which artificially broaden the distribution.

For the grAl samples we observe a fundamentally different behavior of $\std$ with temperature. For sample grAl~A, $\std$ first increases, then decreases starting from $~900\,\mathrm{mK}$ and saturates above 1.3\,K. The width of the measured distributions of sample grAl~B and grAl~C decreases starting from the lowest measurement temperature and remains almost constant above 1\,K. A similar temperature dependence of the width has been reported for different kinds of Josephson junctions in~\cite{Krasnov2005,Kivioja2005,Mannik2005} and was attributed to a diffusive motion of the phase particle through the potential, before the junctions switch to the resistive state. This is referred to as phase diffusion.

\subsection{Phase diffusion}
The diffusive motion of the phase particle in the junction or SQUID potential is a result of multiple consecutive escape and retrapping events. In the phase diffusion regime, a single escape of the particle does not necessarily lead to a switching of the junction to the resistive state. Evidence for phase diffusion can be found in the decrease of the switching current distribution width with increasing temperature. The reduction of the width occurs as escape events at lower bias currents are more likely to be retrapped and thus less likely to lead to a switching event~\cite{Fenton2008}. As the temperature increases, the retrapping probability for the escaped particle also increases and the width further decreases. 

Although a decreasing critical current also leads to a reduced distribution width~\cite{Li2011}, the dependence of $\std$ on $\ic$ in \eref{eq:width} is insufficient to explain the measured curves in \fref{fig:3} (b) and does not result in the saturation of $\std$ at temperatures close to $\tc$. In contrast, the measurements are qualitatively similar to the results obtained by Fenton and Warburton~\cite{Fenton2008} from Monte Carlo simulations on moderately damped Josephson junctions (quality factor~$Q=7$), assuming a finite retrapping probability in the regime where the escape probability is non-zero. The simulations not only describe the decrease of the width with increasing temperature but also reproduce the relatively constant $\std$ for temperatures close to $\tc$.

\begin{figure*}
\begin{center}
\includegraphics[scale=1]{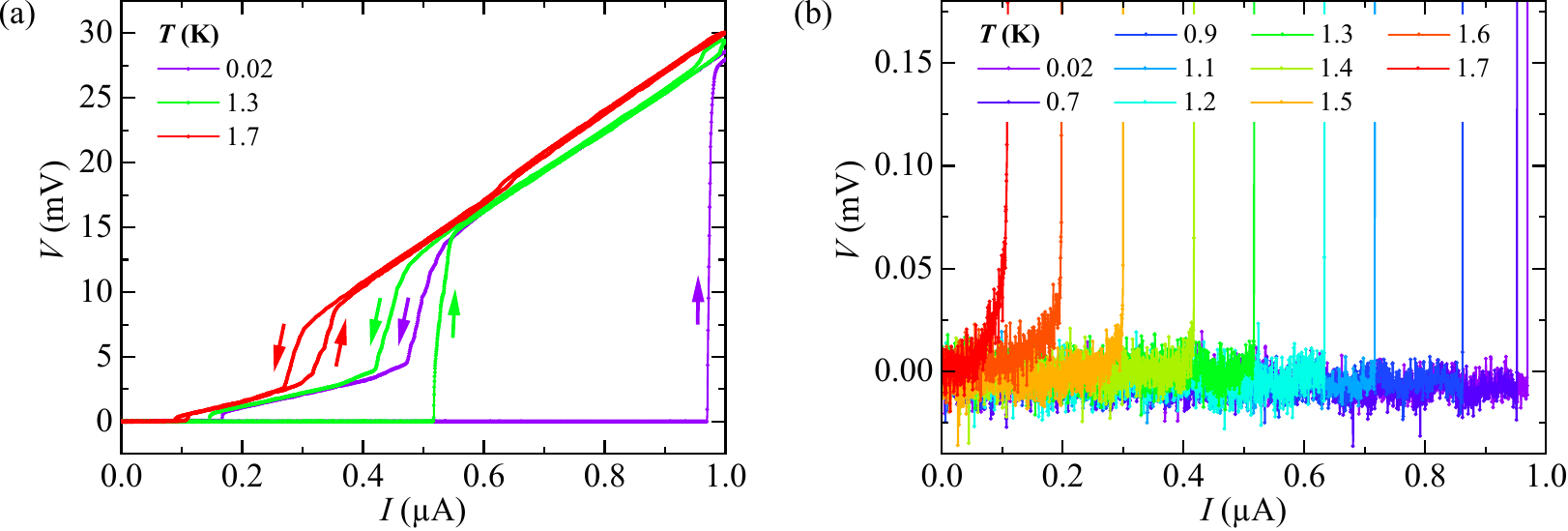}
\caption{(a) Measured I-V curves of sample grAl~B at selected temperatures. The arrows mark the sweeping direction of the respective curve. After the initial switching of the junctions, the behavior is dominated by ohmic heating. (b) Zoom into the \mbox{I-V} characteristics at temperatures ranging from base temperature to 1.7\,K. At temperatures above 1.3\,K, we detect an increase in the voltage before the jump of the SQUID to the resistive state, indicative of phase diffusion.}
\label{fig:4}
\end{center}
\end{figure*}

In \fref{fig:4}~(a), we depict selected \mbox{I-V} curves of sample grAl~B, measured at three different temperatures. After the switching to the resistive state occurred, heat propagation turns large parts of the sample, including the leads, into the normal state, as can be deduced from the measured resistance \mbox{($\sim31\,\mathrm{k}\Omega$)}, which is much larger than the estimated normal state resistance of the geometric constrictions \mbox{($\sim4\,\mathrm{k}\Omega$)}. After the maximum bias current is reached, the current is continuously decreased to zero. Several steps in the down sweep indicate that different parts of the sample progressively become superconducting again. As a consequence, the measured current at which the voltage drops back to zero is not the intrinsic retrapping current of the SQUID, which is determined by the damping of the phase particle, but is rather defined by the thermalization of the circuit. This is also confirmed by the fact that the measured retrapping current is independent of the applied magnetic field (not shown)~\cite{Angers2008}.

In addition to the data presented in \fref{fig:3}~(b), further evidence for the presence of phase diffusion in the higher resistivity grAl films can be found in the zoom-in of the measured \mbox{I-V} curves shown in \fref{fig:4}~(b). Phase slips, individual jumps of the superconducting phase by $2\pi$, that do not lead to a switching event, only slowly change the phase across the junction and are therefore not detected as a sharp step in the \mbox{I-V} characteristic. If the phase slip rate is sufficiently high before the switching of the junction, a small but continuous increase of the voltage can be detected. The effect was previously reported, among others, by Sahu \textit{et al.} in a superconducting nanowire junction~\cite{Sahu2009} and is also observed in our measurements. At temperatures above 1.2 K, we measure a small increase of the voltage, before the SQUID switches to the resistive state. From the maximum voltage measured right before the switching event, which is roughly 50\,$\upmu$V at 1.7 K, a phase slip rate of \mbox{$\sim 2\times10^{10}$} phase slips per second can be calculated. This is in good agreement with the plasma frequency of 70\,GHz predicted and measured for a similar grAl film with a resistivity of 3000\,$\upmu\Omega$\,cm~\cite{Maleeva2018}. The lack of detectable voltage tails at lower temperatures can be explained by a reduced phase slip rate, which only leads to a voltage drop smaller than the experimental noise. The reduction of phase slips might be due to a smaller retrapping probability and hence switching at smaller bias currents.

The experimental results obtained from the switching current measurements suggest the presence of phase diffusion in grAl micro-SQUIDs. For high resistivity films, phase diffusion is present starting from the the lowest measurement temperature, while sample grAl~A with a lower normal state resistivity shows a crossover between a regime without and with phase diffusion. The absence of switching events from excited states in the SQUID modulation is in accordance with the observed phase diffusion. As soon as multiple escapes are necessary to trigger the switching of the SQUID, the switching event always occurs from the ground state of the SQUID potential. Therefore, no excited flux states are visible for sample grAl~B and grAl~C and they disappear at the crossover temperature for sample grAl~A, although $\bl\gg 1$ for all grAl SQUIDs.

We suggest two possible explanations for the onset of phase diffusion in the high resistivity samples. One reason might be a change of the quality factor~$Q=\omega_\mathrm{p}RC$ of our SQUIDs, where $R$ and $C$ account for dissipation and shunt capacitance of the SQUID, respectively. From measurements presented in \cite{LevyBertrand2019} we know that the plasma frequency of the high resistivity samples is in the range of 50-75\,GHz, well below the spectroscopic gap of grAl. Therefore, the fact that these films show increased dissipation compared to sample grAl~A and pure Al, for which the plasma frequency is above the spectroscopic gap, is intriguing, as quasiparticle excitations should be diminished for the high resistivity samples. However, the value of $R$ (in the RCSJ model) could potentially be strongly influenced by the increased amount of oxide in samples grAl B and C. Concretely, at the plasma frequency the quality factor will be susceptible to the intrinsic loss tangent of the aluminum oxide in between the Al grains. Precise information about the value of $Q$ and the origin of increased dissipation requires further investigations, such as direct microwave spectroscopy at the plasma frequency or local scanning microscopy. The other possible reason for the observed phase diffusion is an additional damping mechanism at frequencies comparable to $\omega_\mathrm{p}$, originating from strongly coupled and dissipative environmental modes (such as box modes). The increased density of plasmon modes in high resistive samples already reported in \cite{Maleeva2018} could increase the coupling to these environmental modes.

\section{Conclusion}
We measured the switching currents of grAl micro-SQUIDs to investigate the phase dynamics in constrictions made of grAl thin films of various normal-state resistivities. The switching current modulation with applied magnetic field is reduced and shows a triangular shape, due to the large kinetic inductance in the SQUID loop. The measured temperature dependence of the switching current distribution widths suggests the presence of phase diffusion of the SQUID junctions for high resistivity films. In these samples, a measured increase of the voltage in the \mbox{I-V} characteristic of the SQUID before switching to the resistive state further supports the presence of individual phase slips, leading to phase diffusion.  \\
\\
Our results might indicate the presence of increased dissipation at frequencies comparable to the plasma frequency for the most resistive films, which could originate either from intrinsic losses in the aluminum oxide barriers between the Al grains, or from stronger coupling to dissipative environmental modes. These films are particularly attractive for high impedance quantum electronics~\cite{Grunhaupt2018_2}. Determining the exact mechanism responsible for phase diffusion requires additional work, such as high frequency microwave spectroscopy~\cite{Dupre2017}, local probing using scanning tunneling microscopy or terahertz reflectometry~\cite{Pracht2013}.

\ack
We thank A Quintilla for performing the electron beam lithography of our samples. Facilities use was supported by the KIT Nanostructure Service Laboratory (NSL). We acknowledge support from the Initiative and Networking Fund of the Helmholtz Association, within the Helmholtz Future Project Scalable solid state quantum computing. KB is grateful to the Helmholtz Foundation IVF Solid State Spin Quantum Computing project. IMP acknowledges the support of the Alexander von Humboldt foundation in the framework of a Sofja Kovalevskaja award endowed by the German Federal Ministry of Education and Research. WW acknowledges funding from the Alexander von Humboldt foundation.

\appendix
\section*{Appendix}
\appendix
\setcounter{section}{1}

\begin{figure*}
\begin{center}
\includegraphics[scale=1]{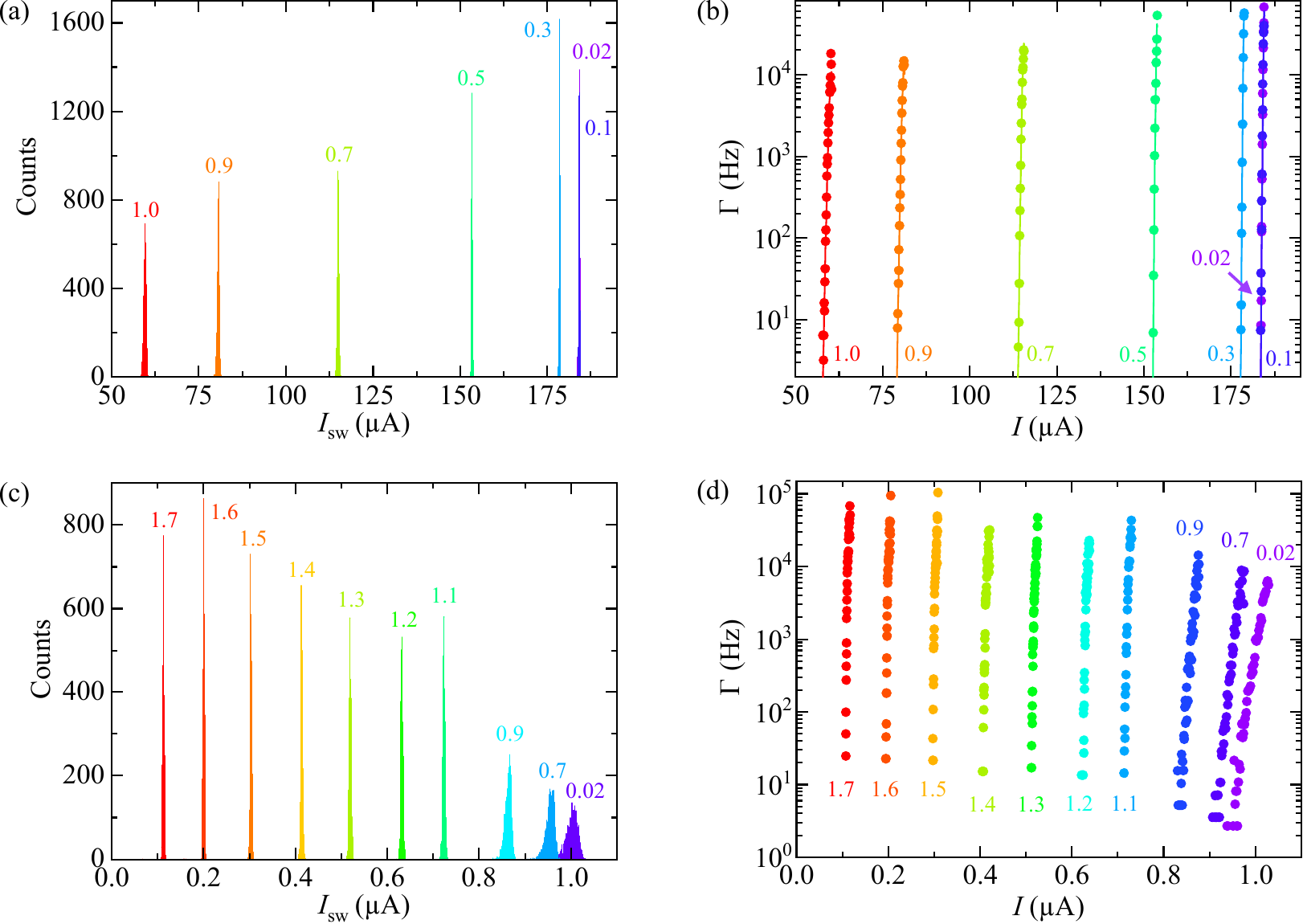}
\caption{(a) Measured switching current histograms of the Al SQUID. (b) Dots show the escape rates extracted from the measurements in (a). Lines are fits to the data according to \eref{eq:appendix}, using a sinusoidal CPR. (c) and (d) Same as (a) and (b) for sample grAl~B. As only multiple escapes lead to a switching event of sample grAl~B, the data in (d) cannot be fitted with \eref{eq:appendix}.}
\label{fig:appendix}
\end{center}
\end{figure*}

In \fref{fig:appendix} we show the histograms and derived escape rates~\cite{Wallraff2003} for the Al SQUID and for sample grAl~B. The histograms illustrate the dependence of $\std$ on temperature and the large $\std/\mean$-ratio of the grAl samples in comparison to the Al SQUID.

The escape rate in the RCSJ model is given by
\begin{eqnarray}
\Gamma(I,T) = \Gamma_0\left(1-I/\ic\right)^\beta\exp\left(-\frac{U_0\left(1-I/\ic\right)^\alpha}{\kb T}\right),\label{eq:appendix}
\end{eqnarray}
where $\Gamma_0$ is the attempt frequency and $U_0$ the potential barrier at zero applied bias current, both depending on $T$. The exponents $\alpha$ and $\beta$ depend on the current phase relation (CPR) of the junction. Note that $\alpha$ here is the same parameter as in \eref{eq:width}. Fitting the extracted escape rates with different sets of $\alpha$ and $\beta$ representing different CPRs and evaluating the quality of the fit can indicate the junction's current phase relation~\cite{Aref2012}. We performed fits of the Al SQUID data for both a sinusoidal CPR ($\alpha=3/2$ and $\beta=1/4$), which we would expect for tunnel junctions and constrictions with a coherence length much larger than the constriction length, and for a sawtooth like CPR ($\alpha=1$ and $\beta=0$), which is expected for constriction junctions with $\xi$ much smaller than the constriction length~\cite{Likharev1979}. We find good agreement only for the sinusoidal CPR, which is expected for the Al SQUID due to the relatively long coherence length, as already discussed in the main text. In the case of the grAl SQUID, we cannot apply \eref{eq:appendix} to fit the data, as a single escape of the phase particle does not lead to a switching event of the SQUID due to phase diffusion. The proper description of the derived escape rates would require simulations, including detailed knowledge about the intrinsic properties of the junctions, such as the magnitude of damping or the intrinsic retrapping current, which are both not accessible in our current experiments.

\section*{References}
\bibliographystyle{iopart-num}
\bibliography{library}

\end{document}